\documentstyle[12pt,epsfig,axodraw,a4]{article} 
\newcommand{\vs}{\vspace{-0.25cm}}
\begin{document} 

\begin{center}
{\Large{\bf Spin-isospin stability of nuclear matter}}

\bigskip

N. Kaiser\\

\medskip

{\small Physik-Department T39, Technische Universit\"{a}t M\"{u}nchen,
    D-85747 Garching, Germany}

\end{center}

\medskip

\begin{abstract}
We calculate the density-dependent spin-isospin asymmetry energy $J(k_f)$ of
nuclear matter in the three-loop approximation of chiral perturbation theory. 
The interaction contributions to $J(k_f)$ originate from one-pion exchange, 
iterated one-pion exchange, and irreducible two-pion exchange with no, single, 
and double virtual $\Delta$-isobar excitation. We find that the approximation 
to $1\pi$-exchange and iterated $1\pi$-exchange terms (which leads already to 
a good nuclear matter equation of state by adjusting an emerging 
contact-term) is spin-isospin stable, since $J(k_{f0})\simeq 24\,{\rm MeV}>0$. 
The inclusion of the chiral $\pi N\Delta$-dynamics, necessary in order to
guarantee the spin-stability of nuclear matter, keeps this property intact. The
corresponding spin-isospin asymmetry energy $J(k_f)$ stays positive even for 
extreme values of an undetermined short-distance parameter $J_5$ (whose 
possible range we estimate from realistic NN-potentials). The largest positive 
contribution to $J(k_f)$ (a term linear in density) comes from a two-body 
contact-term with its strength fitted to the empirical nuclear matter 
saturation point. 
\end{abstract}

\bigskip



\vspace{1cm}

In recent years a novel approach to the nuclear matter problem has emerged. 
Its key element is a separation of long- and short-distance dynamics and an 
ordering scheme in powers of small momenta. At nuclear matter saturation 
density $\rho_0 \simeq 0.16\,$fm$^{-3}$ the Fermi momentum $k_{f0}$ and the 
pion mass $m_\pi$ are comparable scales ($k_{f0}\simeq 2 m_\pi$), and 
therefore pions must be included as explicit degrees of freedom in the
description of the nuclear many-body dynamics. The contributions to the energy
per particle $\bar E(k_f)$ of isospin-symmetric (spin-saturated) nuclear 
matter as they originate from chiral pion-nucleon dynamics have been computed 
up to three-loop order in Refs.\cite{lutz,nucmat}. Both calculations are able
to reproduce the empirical saturation point of nuclear matter by adjusting one
single parameter (either a contact-coupling $g_0+g_1 \simeq 3.23$ \cite{lutz} 
or a cutoff scale $\Lambda \simeq 0.65\,$GeV \cite{nucmat}) related to 
unresolved short-distance dynamics.\footnote{Fitting a cutoff scale, as done
in Ref.\cite{nucmat}, must be viewed as a short-term intermediate step before 
an eventual full effective field theory calculation. Cutoff independence of
physical observables is in fact a primary goal of effective field theory.} The 
basic mechanism for saturation in this approach is a repulsive contribution to
the energy per particle $\bar E(k_f)$ generated by Pauli-blocking in second
order (iterated) pion-exchange. As outlined in Sec. 2.5 of Ref.\cite{nucmat}
this mechanism becomes particularly transparent by taking the chiral limit
$m_\pi = 0$. In that case the interaction contributions to  $\bar E(k_f)$ are
completely summarized by an attractive $k_f^3$-term and a repulsive
$k_f^4$-term where the parameter-free prediction for the coefficient of the
latter is very close  to the one extracted from a realistic nuclear matter
equation of state.   

In a recent work \cite{deltamat} we have extended the chiral approach to 
nuclear matter by including systematically the effects from $2\pi$-exchange 
with virtual $\Delta(1232)$-isobar excitation. The physical motivation for 
such an extension is threefold. First, the spin-isospin-3/2 $\Delta(1232)
$-resonance is the most prominent feature of low-energy $\pi N$-scattering. 
Secondly, it is well known that $2\pi$-exchange between nucleons with 
excitation of virtual $\Delta$-isobars generates the needed isoscalar central 
NN-attraction \cite{gerst} which in phenomenological one-boson exchange models 
is often simulated by a fictitious  scalar ''$\sigma$''-meson exchange. 
Thirdly, the delta-nucleon mass splitting $\Delta = 293\,$MeV is of the same 
size as the Fermi momentum $k_{f0} \simeq 2m_\pi$ at nuclear matter saturation 
density and therefore pions and $\Delta$-isobars should both be treated as 
explicit degrees of freedom. A large variety of nuclear matter properties has 
been investigated in this extended framework in Ref.\cite{deltamat}. It has 
been found that the inclusion of the chiral $\pi N \Delta$-dynamics is able to 
remove most of the shortcomings of previous chiral calculations of nuclear 
matter \cite{nucmat,pot,liquidgas,lutzcontra}. However, there remain open 
questions concerning the role of yet higher orders in the small momentum
expansion and its ''convergence''. The relation of the fitted short-distance 
parameters \cite{nucmat} to those of few-nucleon systems is not clear at this 
moment. Also, a rigorous power counting that justifies the perturbative chiral 
expansion for nuclear matter has not yet been formulated. Recent work by 
Bogner et al.\,\cite{bogner} based on the universal low-momentum NN-potential 
$V_{\rm low-k}$ may open interesting perspectives in this direction.    

Irrespective of such foundational questions it is also necessary to check 
various stability conditions for nuclear matter in the chiral framework. In a
recent paper \cite{spinstab} we have analyzed spin-stability. It turned that
the inclusion of the chiral $\pi N\Delta$-dynamics is essential in order to 
guarantee the spin-stability of isospin-symmetric nuclear matter. The 
truncation to fourth order terms in the small momentum expansion with
interaction contributions only from $1\pi$- and iterated $1\pi$-exchange is 
spin-unstable \cite{spinstab}. This statement holds independently of the 
regularization scheme if the contact-terms (generating contributions linear in
the nucleon density) are consistent with the empirical nuclear matter bulk
properties: $\bar E(k_{f0})\simeq -16\,$MeV and $A(k_{f0}) \simeq 34\,$MeV. 
Now, since a nucleon possesses four internal spin and isospin degrees of
freedom one can prepare nuclear matter also in a spin-isospin mixed asymmetric
configuration. The stability of nuclear matter against such correlated
spin-isospin deformations is the subject of the present paper. For recent work
on generalized symmetry  energy coefficients in the context of 
phenomenological Skyrme forces, also see Ref.\cite{braghin}. Analogous earlier 
studies within Brueckner theory using the Reid soft-core NN-potential can be
found in Ref.\cite{dabr} .  

Let us begin with defining the spin-isospin asymmetry energy $J(k_f)$ of 
(infinite) nuclear matter. Consider this many-nucleon system in a state where 
the equal densities of the spin-up protons $(p\!\!\uparrow)$ and the spin-down 
neutrons $(n\!\!\downarrow)$ have an excess over the equal densities of the 
spin-down protons $(p\!\!\downarrow)$ and the spin-up neutrons $(n\!\!\uparrow
)$. With the help of the spin- and isospin-projection operators,
$(1\pm\sigma_3)/2$ and $(1\pm \tau_3)/2$, such a spin-isospin mixed
asymmetric configuration is realized by the substitution:  
\begin{equation} \theta(k_f-|\vec p\,|) \quad \to \quad {1+\sigma_3 \tau_3 
\over 2}\, \theta(k_+-|\vec p\,|) +{1-\sigma_3 \tau_3 \over 2}\, \theta(k_-
- |\vec p\,|)\,, \end{equation}  
in the medium insertion.\footnote{Medium insertion is a technical notation for
the difference between the in-medium and vacuum nucleon propagator 
\cite{nucmat}. Effectively, it sums hole-propagation and the absence of
particle-propagation below the Fermi surface $|\vec p\,| < k_f$.} Here, $k_+ = 
k_f( 1+ \epsilon )^{1/3}$ and $k_- = k_f( 1- \epsilon)^{1/3}$ (with $\epsilon$
a small parameter) are different Fermi momenta, chosen such that the total
nucleon density $\rho = (k^3_+ +k^3_-)/3\pi^2 = 2k_f^3/3\pi^2$ stays
constant. Note that Eq.(1) describes a rather peculiar asymmetric
configuration of nuclear matter with equal densities of protons, neutrons,
spin-up states and spin-down states: $\rho_p = \rho_n = \rho_\uparrow
=\rho_\downarrow = k_f^3/3\pi^2$. The expansion of the energy per particle of
spin-isospin polarized nuclear matter: 
\begin{equation} \bar E(k_+,k_-)_{\sigma\tau-{\rm pol}}= \bar E(k_f) + 
\epsilon^2\, J(k_f)+ {\cal O}(\epsilon^4) \,, \qquad  k_\pm = k_f(1\pm
\epsilon)^{1/3} \,,  \end{equation}
defines the spin-isospin asymmetry energy $J(k_f)$. The obvious criterion for 
the spin-isospin stability of nuclear matter is then the positivity of the 
spin-isospin asymmetry energy: $J(k_f)>0$. The energy per particle at fixed
nucleon density $\rho$ must take on its absolute minimum value in the
spin- and isospin-saturated configuration.   

The first contribution to the spin-isospin asymmetry energy $J(k_f)$ comes 
from the kinetic energy $\sqrt{M^2+p^2}-M$ of a non-interacting relativistic
Fermi gas of nucleons: 
\begin{equation} J(k_f)={k_f^2\over 6 M}-{k_f^4\over 12 M^3}\,,\end{equation}
with $M=939\,$MeV the (average) nucleon mass. The next term in this series, 
$k_f^6/16M^5$, is negligibly small at the densities of interest. 

\begin{figure}
\begin{center}
\includegraphics[scale=1.,clip]{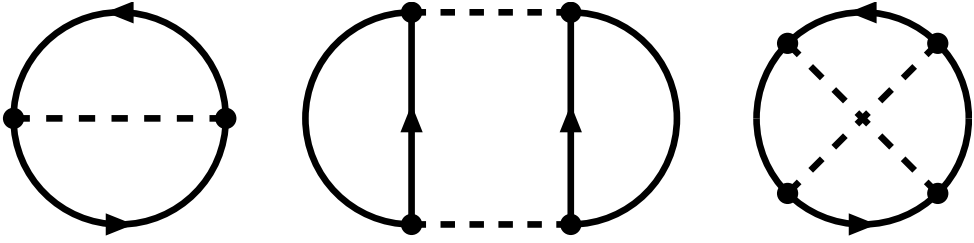}
\end{center}\vspace{-0.4cm}
\caption{The two-loop one-pion exchange Fock diagram and the three-loop
iterated one-pion exchange Hartree and Fock diagrams. The combinatoric factors
of these diagrams are $1/2$, $1/4$ and $1/4$, in the order shown.}
\end{figure}

Next, we come to interaction contributions to $J(k_f)$. The closed in-medium 
diagrams related to one-pion exchange (Fock diagram) and iterated one-pion
exchange (Hartree and Fock diagrams) are shown in Fig.\,1. Differences in 
comparison to the calculation of the energy per particle $\bar E(k_f)$ in
Ref.\cite{nucmat} occur only with respect to the factors emerging from the
spin and isospin traces over closed nucleon lines and the radii $k_\pm=k_f(1
\pm \epsilon)^{1/3}$ of the Fermi spheres to be integrated over. After some
analytical calculation we find the following contribution to the spin-isospin 
asymmetry energy $J(k_f)$ from the $1\pi$-exchange Fock diagram in Fig.\,1 
(including its relativistic $1/M^2$-correction):  
\begin{eqnarray} J(k_f) &=& {g_A^2m_\pi^3 \over(4\pi f_\pi)^2} 
\bigg\{{u^3\over 9}-{u\over 2}+\bigg( {2u\over 9}+{1\over 8u}\bigg)
\ln(1+4u^2) \nonumber \\ && +{m_\pi^2\over M^2} \bigg[{19u^3\over 18}- {4u^5
\over 9}-{u^2\over 2} \arctan 2u -{u \over 72} (1+18u^2)\ln(1+4u^2)\bigg] 
\bigg\}\,. \end{eqnarray} 
Here, we have introduced the abbreviation $u=k_f/m_\pi$ where $m_\pi 
=135\,$MeV stands for the (neutral) pion mass. As usual $f_\pi = 92.4\,$MeV 
denotes the weak pion decay constant and we choose the value $g_A=1.3$ of the
nucleon axial-vector coupling constant in order to have a pion-nucleon 
coupling constant of $g_{\pi N} = g_A M/f_\pi = 13.2$. In the 
second and third diagram in Fig.\,1 the $1\pi$-exchange interaction is 
iterated (once) with itself. These second order diagrams carry the large scale
enhancement factor $M$ (the nucleon mass). It stems from an energy denominator
that is equal to a difference of small nucleon kinetic energies. With a medium
insertion at each of two equally oriented nucleon propagators we obtain from
the three-loop Hartree diagram in Fig.\,1 the following contribution to the
spin-isospin asymmetry energy:  
\begin{equation} J(k_f)= {\pi g_A^4 M m_\pi^4 \over 6(4\pi f_\pi)^4}\bigg\{
\bigg( 15u +{7\over 2u}\bigg) \ln(1+4u^2)-14u -16u^2 \arctan 2u \bigg\} \,. 
\end{equation}
The right Fock diagram of iterated $1\pi$-exchange (see Fig.\,1) with two
medium insertions on non-neighboring nucleon propagators gives rise on the 
other hand to a contribution to the spin-isospin asymmetry energy of the form:
\begin{eqnarray} J(k_f) &=& {\pi g_A^4 M m_\pi^4 \over 9(4\pi f_\pi)^4}\bigg\{
{21u\over 5} -{64u^3 \over 15}-\bigg(9 +16u^2 +{64 u^4\over 15}\bigg)\arctan u 
\nonumber \\  &&  +\bigg( {33 \over 10 u}+{14u \over 3} \bigg) \ln(1+u^2) 
-\bigg( {3\over u}+2u\bigg) \ln(1+4u^2)\nonumber \\ && + (9-4u^2)\arctan 2u  + 
(9+4u^2) \int_0^u \!dx\, { \arctan x-\arctan 2x \over u(1+2x^2)} \bigg\}\,.  
\end{eqnarray}
This expression does not include the contribution of a linear divergence
$\int^\infty_0 dl\,1$ of the momentum-space loop integral. In dimensional 
regularization such a linear divergence is set to zero, whereas in cut-off 
regularization it is equal to a momentum space cut-off $\Lambda$. The 
additional term specific for cut-off regularization will be given in Eq.(13). 
An in-medium diagram with three medium insertions represents Pauli-blocking 
effects in intermediate NN-states induced by the filled Fermi sea of nucleons. 
The unequal filling of the $(p\!\!\uparrow, n\!\!\downarrow)$ and $(p\!\!
\downarrow, n\!\!\uparrow)$ Fermi seas shows its consequences in the 
spin-isospin asymmetry energy. After some extensive algebraic manipulations we
end up with the following double-integral representation of the contribution 
to the spin-isospin asymmetry energy $J(k_f)$ from the Hartree diagram in
Fig.\,1 with three medium insertions:  
\begin{eqnarray} J(k_f)&=&{g_A^4 M m_\pi^4 \over (4\pi f_\pi)^4 u^3} \int_0^u
\! dx \,x^2 \int_{-1}^1 \! dy \bigg\{ \bigg[{2uxy(3u^2-5x^2y^2) \over (u^2-
x^2y^2)} -(u^2+5x^2y^2)H\bigg] \nonumber \\ && \times \bigg[{2s^2+s^4
\over 1+s^2}-2\ln(1+s^2) \bigg]  +{4u^2 H \,s^5(8s'-9s) \over 9(1+s^2)^2}+ 
\Big[2uxy+(u^2-x^2y^2) H\Big] \nonumber \\ && \times \Big[(5+s^2)(9s^2-16s s' 
+16 s'^2)+8s(1+s^2)(2s''-10s'+9s)\Big] {s^4\over 9(1+s^2)^3} 
\bigg\} \,, \end{eqnarray}
where we have introduced several auxiliary functions:
\begin{equation} H = \ln{u+x y\over u- xy}\,, \qquad s= xy
+\sqrt{u^2-x^2+x^2y^2}\,, \qquad s' = u \,{\partial s \over \partial u} \,, 
\qquad s'' = u^2 \, {\partial^2 s \over \partial u^2} \,. \end{equation}
Note that Eq.(7) stems from a nine-dimensional principal-value integral over 
the product of three Fermi spheres of varying radii $k_\pm= k_f(1\pm \epsilon
)^{1/3}$ which has been differentiated twice with respect to $\epsilon$ at 
$\epsilon=0$. Of similar structure is the contribution to $J(k_f)$ from the  
iterated $1\pi$-exchange Fock diagram with three medium insertions. Because of 
the two different pion propagators in the Fock diagram one ends up (partially) 
with a triple-integral representation for its contribution to the spin-isospin
asymmetry energy:       
\begin{eqnarray}  J(k_f) &=&{g_A^4Mm_\pi^4\over 72(4\pi f_\pi)^4 u^3}\int_0^u\!
dx\Bigg\{G(9G_{20}+2G_{11}+9G_{02}-16G_{01}-9G) \nonumber \\ && +9G^2_{10} 
+2G_{01}G_{10}-5G_{01}^2 +4x^2 \int_{-1}^1\!dy \int_{-1}^1 \!dz {yz \,\theta
(y^2+z^2-1) \over |yz| \sqrt{y^2+z^2-1}} \nonumber \\ && \times \bigg[{2s^3t^3 
(16s't-9st-12s't') \over (1+s^2)(1+t^2)}+ {s^2[t^2-\ln(1+t^2)]\over (1+s^2)^2} 
\nonumber \\ && \times \Big[(3+s^2)(16ss'-9s^2-16s'^2)+4s (1+s^2)(12s'-9s-4s'')
\Big] \bigg]\Bigg\} \,.  \end{eqnarray}
Here, we have split into factorizable and non-factorizable parts. These two 
pieces are distinguished by whether the (remaining) nucleon propagator in the 
three-loop Fock diagram can be canceled or not by terms from the product of 
$\pi N$-interaction vertices. The factorizable terms can be expressed through 
the auxiliary function:   
\begin{equation} G = u(1+u^2+x^2) -{1\over 4x}\big[1+(u+x)^2\big] \big[1+
(u-x)^2\big] \ln{1+(u+x)^2\over 1+(u-x)^2 } \,,   \end{equation}
and its partial derivatives for which we have introduced a (short-hand) 
double-index notation: 
\begin{equation}  G_{ij} = x^i u^j  {\partial^{i+j}G \over \partial x^i 
\partial u^j} \,, \quad 1\leq i+j \leq 2\,. \end{equation}
For the presentation of the nonfactorizable terms one needs also copies of the
quantities $s$ and $s'$ defined in Eq.(8) which depend (instead of $y$) on
another directional cosine $z$:  
\begin{equation} t= xz +\sqrt{u^2-x^2+x^2z^2}\,, \qquad t' = u\, {\partial t 
\over \partial u} \,. \end{equation}
In the chiral limit $m_\pi = 0$ the fourth order contributions in Eqs.(5-9) 
sum up to a negative  $k_f^4$-term of the form: $J(k_f)|_{m_\pi =0} = -(g_Ak_f
/4\pi f_\pi)^4 (M/405) (32\pi^2+741+1848 \ln2)$. Finally, we give the 
expression for the linear divergence specific to cut-off regularization:   
\begin{equation} J(k_f) =  {10 g_A^4 M \Lambda \over 3(4\pi f_\pi)^4}\, k_f^3 
\,, \end{equation}
to which only the iterated $1\pi$-exchange Fock diagram (with two medium 
insertions) has contributed. In the case of the Hartree diagram the linear
divergence drops out after taking the second derivative with respect to 
$\epsilon$. One observes that the term in Eq.(13) is just $-1/3$ of the 
corresponding contribution to the energy per particle $\bar E(k_f)$ (see 
Eq.(15) in Ref.\cite{nucmat}). In this context it is interesting to note that 
for terms linear in density $\rho$ the relation $3J(k_f)_{\rm lin}=-\bar 
E(k_f)_{\rm lin}$ holds generally. It is a consequence of the spin-isospin 
structure $3-\vec\sigma_1 \cdot \vec \sigma_2\, \vec\tau_1 \cdot \vec \tau_2$ 
of a Fierz-antisymmetric NN-contact interaction (see e.g. Eq.(35) in 
Ref.\cite{nnpap}).

\begin{figure}
\begin{center}
\includegraphics[scale=0.65,clip]{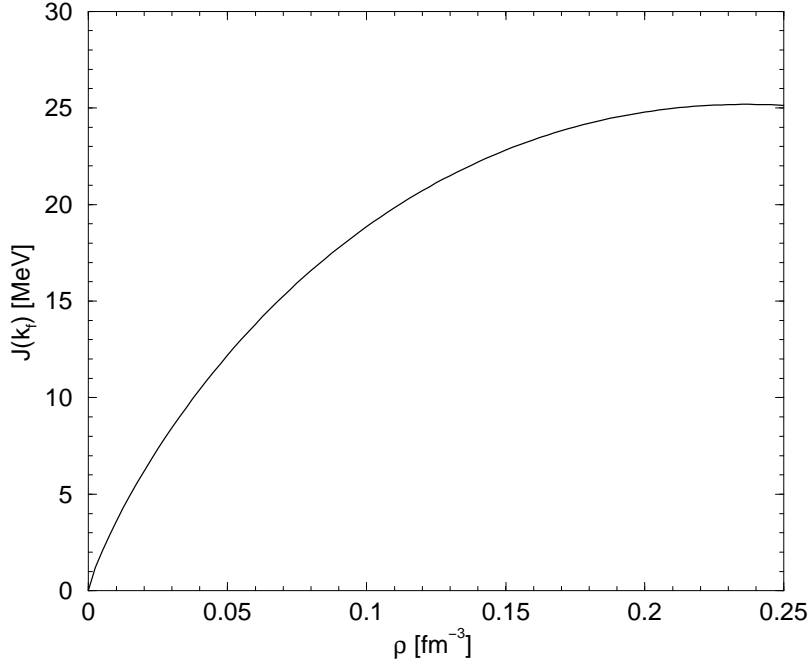}
\end{center}\vspace{-0.8cm}
\caption{The spin-isospin asymmetry energy $J(k_f)$ of nuclear matter versus 
the nucleon density $\rho= 2k_f^3/3\pi^2$. The solid line shows the result of a
calculation up to fourth order in small momenta including $1\pi$-exchange
and iterated $1\pi$-exchange. The cut-off scale $\Lambda = 0.61\,$GeV has
been adjusted to the saturation point: $\rho_0 = 0.173\,$fm$^{-3}$, $\bar
E(k_{f0})= -15.3\,$MeV. The positive values of $J(k_f)$ indicate the
spin-isospin stability of nuclear matter in this approximation.}  
\end{figure}

Now we can turn to numerical results. In Fig.\,2 we show the spin-isospin 
asymmetry energy $J(k_f)$ of nuclear matter as a function the nucleon density
$\rho=2k_f^3/3\pi^2$. The solid line corresponds to a calculation up to fourth
order in small momenta. It includes besides the kinetic energy term Eq.(3) the
contributions from static $1\pi$-exchange and iterated $1\pi$-exchange. For 
reasons of consistency we have dropped the small relativistic 
$1/M^2$-correction in Eq.(4) since it is of fifth order in the small momenta  
$k_f$ and $m_\pi$. The cut-off scale $\Lambda =0.61\,$GeV has been adjusted 
\footnote{This provisional procedure introduces a model-dependence that lies 
outside effective field theory.} to the nuclear matter saturation point 
$\rho_0= 0.173\,$fm$^{-3}$ and $\bar E(k_{f0})= -15.3\,$MeV. The resulting
value of the nuclear matter compressibility $K = k^2_{f0}\bar E''(k_{f0})= 
252\,$MeV is consistent with a recent extrapolation from giant monopole 
resonances of heavy nuclei \cite{dario}, which gave $K = (260\pm 10)\,$MeV. 
One can read off from Fig.\,2 a positive value of the spin-isospin asymmetry 
energy at saturation density: $J(k_{f0}) =J(2m_\pi) = 23.9\,$MeV. It indicates 
the spin-isospin stability of nuclear matter in this approximation. The 
largest positive contribution to $J(2m_\pi) = 23.9\,$MeV comes from the term, 
Eq.(13), linear in density and amounts to $59.2\,$MeV at saturation density 
$k_{f0} = 2m_\pi$. Compared to that the largest negative contribution is 
$-32.8\,$MeV and it stems from the iterated $1\pi$-exchange Fock diagram with 
two medium insertions, Eq.(6). The remaining numerically smaller contributions
cancel each other to a large extent. It must however be stressed that at this 
level of approximation, with interaction terms only from $1\pi$-exchange and 
iterated $1\pi$-exchange, nuclear matter is spin-unstable \cite{spinstab}. The
inclusion of higher order terms (in particular $2\pi$-exchange with virtual
$\Delta$-isobar excitation) is mandatory in order to achieve spin-stability of
nuclear matter.

\begin{figure}
\begin{center}
\includegraphics[scale=1.,clip]{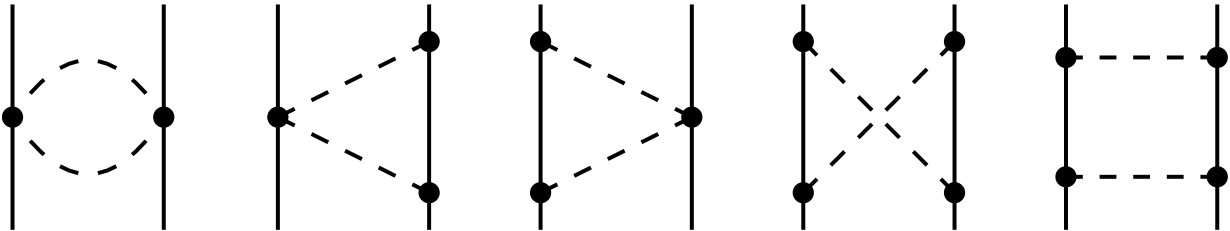}
\end{center}\vspace{-0.4cm}
\caption{One-loop diagrams of irreducible two-pion exchange between nucleons.}
\end{figure}

\begin{figure}
\begin{center}
\includegraphics[scale=0.994,clip]{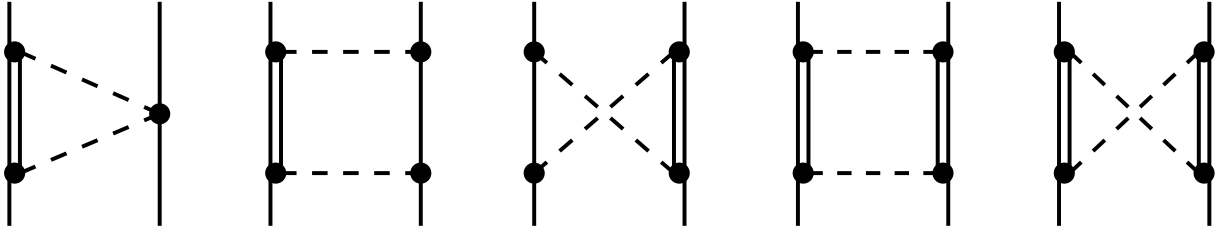}
\end{center}\vspace{-0.4cm}
\caption{One-loop two-pion exchange diagrams with single and double
$\Delta(1232)$-isobar excitation. Diagrams for which the role of both nucleons
is interchanged are not shown.}
\end{figure}

Therefore, we turn now to contributions to $J(k_f)$ of fifth order in the small
momentum expansion. At three-loop order these terms are generated by 
(irreducible) two-pion exchange between nucleons. The corresponding one-loop 
diagrams for elastic NN-scattering are shown in Fig.\,3. Since we are counting
the delta-nucleon mass splitting $\Delta= 293\,$MeV (together with $k_f$ and
$m_\pi$) as a small momentum scale the diagrams with single and double virtual
$\Delta(1232)$-isobar excitation shown in Fig.\,4 belong to the same order. 
By closing the two open nucleon lines of the one-loop diagrams in Figs.\,3,4 
to either two or one ring one gets (in diagrammatic representation) the
Hartree or Fock contribution to the energy density. The Hartree contribution 
to the spin-isospin asymmetry energy $J(k_f)$ vanishes identically because the 
relevant $2\pi$-exchange NN T-matrix in forward direction is spin-independent 
\cite{gerst,nnpap}. The Fock contribution on the other hand is obtained by 
integrating the spin- and isospin-contracted T-matrix over the product of two 
Fermi spheres of radii $k_\pm = k_f(1\pm \epsilon)^{1/3}$. We separate 
regularization dependent short-range contributions to the T-matrix 
(originating from the ultra-violet divergences of the one-loop diagrams in 
Figs.\,3,4) from the unique long-range terms with the help of a 
twice-subtracted dispersion relation. The occurring subtraction constants give
rise to a contribution to the spin-isospin asymmetry energy of the form: 
\begin{equation} J(k_f)= -B_3{k_f^3 \over 3 M^2} +  J_5 {k_f^5 \over
M^4}\,. \end{equation}
The dimensionless parameters $B_3 = -7.99$ has been adjusted in 
Ref.\cite{deltamat} to the saturation minimum $\bar E(k_{f0})=-16\,$MeV. 
Again, we recognize in the first part of Eq.(14) the relation $3J(k_f)_{\rm 
lin}=-\bar E(k_f)_{\rm lin}$ for terms linear in the density $\rho= 2k_f^3/3
\pi^2$. The other subtraction constant $J_5$ in front of the $k_f^5/M^4$-term 
is (a priori) not constrained by any empirical (ground-state) property of 
nuclear matter. The long-range parts of the $2\pi$-exchange (two-body) Fock 
diagrams can be expressed as a dispersion-integral:  
\begin{eqnarray} J(k_f)&=&  {1 \over 6\pi^3} \int_{2m_\pi}^{\infty} \!\!
d\mu  \bigg\{{\rm Im}(3W_C+2\mu^2 V_T +4\mu^2W_T){k_f\over 3}\bigg[ {4k_f^2
\over \mu} -{8k_f^4 \over \mu^3}- \mu \ln\bigg(1+{4k_f^2 \over \mu^2}\bigg) 
\bigg] \quad \nonumber \\ && + {\rm Im}(V_C+3W_C+2\mu^2 V_T+6\mu^2W_T)\bigg[
{\mu k_f\over 2} -{k_f^3 \over \mu} +{8k_f^5 \over 3\mu^3} - {\mu^3 \over
8k_f}\ln \bigg( 1+ {4k_f^2 \over \mu^2} \bigg) \bigg] \bigg\}, \end{eqnarray}
where Im$V_C$, Im$W_C$, Im$V_T$ and Im$W_T$ are the spectral functions of the
isoscalar and isovector central and tensor NN-amplitudes, respectively. 
Explicit expressions of these imaginary parts for the contributions of the 
triangle diagram with single $\Delta$-excitation and the box diagrams with 
single and double $\Delta$-excitation can be easily constructed from the 
analytical formulas given in Sec.\,3 of Ref.\cite{gerst}. The $\mu$- and 
$k_f$-dependent weighting functions in Eq.(15) take care that at low and
moderate densities this spectral-integral is dominated by low invariant
$\pi\pi$-masses $2m_\pi< \mu <1\,$GeV. The contributions to the spin-isospin 
asymmetry energy $J(k_f)$ from irreducible $2\pi$-exchange (with only nucleon 
intermediate states, see Fig.\,3) can also be cast into the form Eq.(15). The 
corresponding non-vanishing spectral functions read \cite{nnpap}: 
\begin{equation} {\rm Im}W_C(i\mu) = {\sqrt{\mu^2-4m_\pi^2} \over 3\pi 
\mu (4f_\pi)^4} \bigg[ 4m_\pi^2(1+4g_A^2-5g_A^4) +\mu^2(23g_A^4-10g_A^2-1) + 
{48 g_A^4 m_\pi^4 \over \mu^2-4m_\pi^2} \bigg] \,, \end{equation}
\begin{equation} {\rm Im}V_T(i\mu) = - {6 g_A^4 \sqrt{\mu^2-4m_\pi^2} \over 
\pi  \mu (4f_\pi)^4}\,. \end{equation}

\begin{figure}
\begin{center}
\includegraphics[scale=1.0,clip]{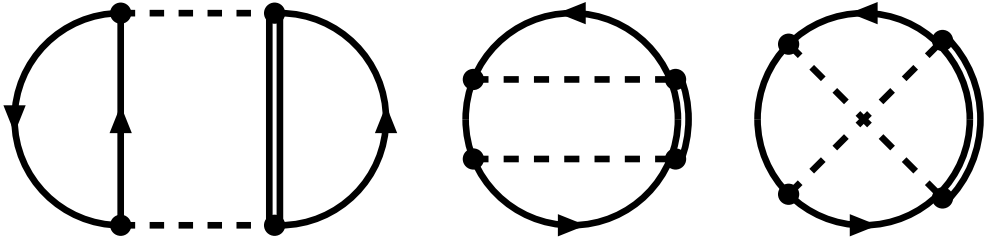}
\end{center}\vspace{-0.4cm}
\caption{Hartree and Fock three-body diagrams related to $2\pi$-exchange with 
single virtual $\Delta$-isobar excitation. They represent interactions between
three nucleons in the Fermi sea. The combinatoric factor is 1 for each 
diagram.} 
\end{figure}

Next, we come to the additional $2\pi$-exchange three-body terms which arise 
from Pauli blocking of intermediate nucleon states (i.e. from the $(1\pm
\sigma_3 \tau_3) \theta(k_\pm -|\vec p\,|)$ terms in the in-medium nucleon 
propagators \cite{nucmat}). The corresponding closed Hartree and Fock diagrams 
with single virtual $\Delta$-excitation are shown in Fig.\,5. The contribution 
of the left three-body Hartree diagram to the spin-isospin asymmetry energy 
$J(k_f)$ has the following analytical form:  
\begin{equation} J(k_f)={g_A^4 m_\pi^6 u^2\over 27\Delta(2\pi f_\pi)^4} \bigg[
\bigg({27\over 4}+8u^2\bigg)\ln(1+4u^2)+2u^4(1-9\zeta) -22u^2-{5u^2\over 
1+4u^2} \bigg] \,. \end{equation}
The delta propagator shows up in this expression merely via the (reciprocal)
mass-splitting $\Delta =293\,$MeV. Furthermore, we have already inserted in 
Eq.(18) the empirically well-satisfied relation $g_{\pi N\Delta} = 3g_{\pi
N}/\sqrt{2}$ for the $\pi N\Delta$-coupling constant. The parameter $\zeta = 
-3/4$ has been introduced in Sec.\,2 of Ref.\cite{deltamat} in order 
to reduce a too strongly repulsive $\rho^2$-term in the energy particle $\bar
E(k_f)$. It controls the strength of a three-nucleon contact interaction
$\sim (\zeta g_A^4/\Delta f_\pi^4) \, (\bar NN)^3$ which has the property that 
it contributes equally but with opposite sign to the energy per particle 
$\bar E(k_f)$ and the spin-isospin asymmetry energy $J(k_f)$. The contribution 
of both three-body Fock diagrams in Fig.\,5 to the spin-isospin asymmetry 
energy $J(k_f)$ can be represented as:   
\begin{eqnarray} J(k_f)&=&{g_A^4 m_\pi^6 \over 108\Delta(4\pi f_\pi)^4u^3}
\int_0^u\!\! dx\Big\{ -4G_{S01}G_{S10}-10G_{S01}^2-18G_{S10}^2\nonumber \\ && +
2 G_S (9G_S+16 G_{S01}-9 G_{S02}-2 G_{S11}-9 G_{S20})- 2G_{T01}G_{T10}
\nonumber \\ && -17G_{T01}^2-9G_{T10}^2  + G_T(9G_T+16 G_{T01}-9G_{T02}
-2G_{T11} -9 G_{T20}) \Big\} \,,\end{eqnarray}
with the two auxiliary functions:
\begin{eqnarray} G_S&=& {4ux \over 3}( 2u^2-3) +4x\Big[
\arctan(u+x)+\arctan(u-x)\Big] \nonumber \\ && + (x^2-u^2-1) \ln{1+(u+x)^2
\over  1+(u-x)^2} \,,\end{eqnarray}
\begin{eqnarray} G_T &=& {ux\over 6}(8u^2+3x^2)-{u\over
2x} (1+u^2)^2  \nonumber \\ && + {1\over 8} \bigg[ {(1+u^2)^3 \over x^2} -x^4 
+(1-3u^2)(1+u^2-x^2)\bigg] \ln{1+(u+x)^2\over  1+(u-x)^2} \,.\end{eqnarray}
The double-indices on $G_S$ and $G_T$ have the same meaning as explained in 
Eq.(11) for the function $G$. 

\begin{figure}
\begin{center}
\includegraphics[scale=0.65]{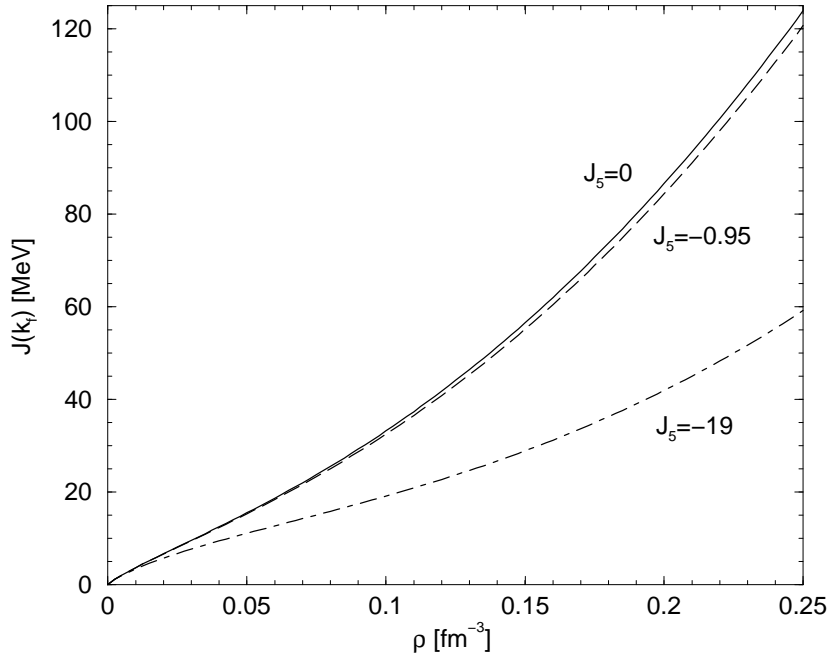}
\end{center}\vspace{-.8cm}
\caption{The spin-isospin asymmetry energy $J(k_f)$ of nuclear matter versus 
the nucleon density $\rho= 2k_f^3/3\pi^2$. In comparison to Fig.\,2 the effects
from $2\pi$-exchange with single and double virtual $\Delta$-isobar excitation
are now included. The solid, dashed, and dashed-dotted curves correspond to the
choices $J_5=0,\,-0.95$ and $-19$ of the short-range parameter $J_5$
introduced in Eq.(14). The positive values of $J(k_f)$ ensure the spin-isospin
stability of nuclear matter.}
\end{figure}

In Fig.\,6 we show again the spin-isospin asymmetry energy $J(k_f)$ of nuclear
matter as a function of the nucleon density $\rho= 2k_f^3/3\pi^2$. The solid 
line includes all the contributions from chiral $1\pi$- and $2\pi$-exchange 
written down in Eqs.(3-9,14-19). The (yet undetermined) short-range parameter 
$J_5$ has been set to zero, $J_5=0$. We note as an aside that the term linear 
in the density and the cut-off $\Lambda$, Eq.(13), is now not counted 
extra since the parameter $B_3=-7.99$ \cite{deltamat} collects all such 
possible terms. Numerically, these two terms linear in density are anyhow 
almost identical. One observes in Fig.\,6 a positive spin-isospin asymmetry 
energy $J(k_f)$ which rises monotonically with the density $\rho$. The 
inclusion of the chiral $\pi N\Delta$-dynamics does therefore not disturb the
spin-isospin stability of nuclear matter. It is also interesting to look at 
numerical values of $J(k_f)$ and their decomposition. At a Fermi momentum of 
$k_f = 2m_\pi$ (corresponding to $\rho = 0.173\,$fm$^{-3}$) the spin-isospin 
asymmetry energy is now $J(2m_\pi)= 69.5\,$MeV (setting $J_5=0$). The most 
significant changes in comparison to the previous fourth order calculation 
come from the two-body Fock and three-body Hartree contributions Eqs.(15,18) 
which amount together to $30.6\,$MeV + $20.7\,$MeV = $51.3\,$MeV. About 
one third thereof (namely $16.6\,$MeV) stems from the three-body contact 
interaction proportional to $\zeta=-3/4$.   

The size of the short-distance parameter $J_5$ in Eq.(14) is still open and 
large negative values could endanger the spin-isospin stability. In order to 
get an estimate of $J_5$ we bring into play the complete set of four-nucleon 
contact-couplings written down in Eqs.(3,4) of Ref.\cite{evgeni}. This set 
represents the most general short-range NN-interaction quadratic in momenta 
and it involves seven low-energy constants $C_1, \dots, C_7$. After computing 
the spin-isospin asymmetry energy $J(k_f)$ from the corresponding
contact-potential in  Hartree-Fock approximation we find:      
\begin{equation} J_5 = {M^4 \over 18\pi^2} (C_2-4C_1) =  {M^4 \over 144\pi^3} 
\Big[3C(^1\!P_1)+C(^3\!P_0) +3C(^3\!P_1)+5 C(^3\!P_2)\Big] \,. \end{equation}  
In the second line of Eq.(22) we have reexpressed the relevant linear 
combination of $C_{1,2}$ through the so-called spectroscopic low-energy
constants which characterize the short-range part of the NN-potential in the
spin-singlet and spin-triplet $S$- and $P$-wave states. In that representation
we obtain from the entries of table IV in Ref.\cite{evgeni} for the three 
NN-potentials \footnote{The short-distance structure of realistic 
NN-potentials and effective field theory could be very different. The idea
here is simply to explore the extreme possible range of $J_5$.} CD-Bonn,
Nijm-II, and AV-18 the numbers: $J_5 = -1.34,\, -0.57$, and $-0.94$. The dashed
line in Fig.\,6 shows the spin-isospin asymmetry energy $J(k_f)$ which results
from taking their average value $J_5= -0.95$. The corresponding reduction of 
the spin-isospin asymmetry energy is negligible. The dashed-dotted curve in 
Fig.\,6 corresponds to the extreme choice $J_5 = -19$. One can see that even 
with such a large negative $J_5$-value the spin-isospin stability of nuclear 
matter remains still preserved. We can therefore conclude that spin-isospin 
stability is a robust property of the chiral approach to nuclear matter (at
least in the three-loop approximation). This is a important finding.

In summary we have investigated in this work the spin-isospin stability of 
nuclear matter in the framework of chiral perturbation theory. For that 
purpose we have calculated the density-dependent spin-isospin asymmetry energy
$J(k_f)$ of nuclear matter to three-loop order. The interaction contributions 
to $J(k_f)$ originate from $1\pi$-exchange, iterated $1\pi$-exchange, and 
(irreducible) $2\pi$-exchange with no, single, and double virtual 
$\Delta$-isobar excitation. We have found that the approximation to $1\pi$- 
and iterated $1\pi$-exchange terms is spin-isospin stable, since $J(k_{f0})
>0$. The inclusion of the chiral $\pi N\Delta$-dynamics (necessary to ensure 
the spin-stability \cite{spinstab} of nuclear matter) keeps this property 
intact. The largest positive contribution to $J(k_f)$ comes from a two-body 
contact interaction with its strength fitted to the empirical nuclear matter 
saturation point.

\end{document}